\documentclass[emulateapj,natbib]{emulateapj}

\begin{document}

\title{Subaru/SCExAO First-Light Direct Imaging of a Young Debris Disk around HD 36546}
\author{Thayne Currie\altaffilmark{1}, Olivier Guyon\altaffilmark{1,2,3}, Motohide Tamura\altaffilmark{2,4},
Tomoyuki Kudo\altaffilmark{1},
Nemanja Jovanovic\altaffilmark{1},
Julien Lozi\altaffilmark{1},
Joshua E. Schlieder\altaffilmark{5},
Timothy D. Brandt\altaffilmark{6},
Jonas Kuhn\altaffilmark{7},
Eugene Serabyn\altaffilmark{8},
Markus Janson\altaffilmark{9},
Joseph Carson\altaffilmark{10},
Tyler Groff\altaffilmark{11},
N. Jeremy Kasdin\altaffilmark{11},
Michael W. McElwain\altaffilmark{12},
Garima Singh\altaffilmark{8},
Taichi Uyama\altaffilmark{13},
Masayuki Kuzuhara\altaffilmark{2}
Eiji Akiyama\altaffilmark{14}, 
Carol Grady\altaffilmark{12,15},
Saeko Hayashi\altaffilmark{1},
Gillian Knapp\altaffilmark{16},
Jung-mi Kwon\altaffilmark{17},
Daehyeon Oh\altaffilmark{18},
John Wisniewski\altaffilmark{19},
Michael Sitko\altaffilmark{20},
Yi Yang\altaffilmark{4}
}
\altaffiltext{1}{Subaru Telescope, National Astronomical Observatory of Japan, National Institutes of Natural Sciences, Hilo, HI, USA}
\altaffiltext{2}{Astrobiology Center, National Institutes of Natural Sciences, 2-21-1 Osawa, Mitaka, Tokyo, Japan} 
\altaffiltext{3}{Steward Observatory, University of Arizona, Tucson, AZ 85721, USA} 
\altaffiltext{4}{Department of Astronomy, Graduate School of Science, The University of Tokyo, 113-0033, Japan}
\altaffiltext{5}{IPAC-NExScI, Mail Code 100-22, Caltech, 1200 E. California Blvd., Pasadena, CA 91125}
\altaffiltext{6}{Astrophysics Department, Institute for Advanced Study, Princeton, NJ}
\altaffiltext{7}{Institute for Astronomy, ETH-Zurich, Wolfgang-Pauli-Str. 27, 8093 Zurich, Switzerland}
\altaffiltext{8}{Jet Propulsion Laboratory, California Institute of Technology, 4800 Oak Grove Dr., Pasadena, CA}
\altaffiltext{9}{Department of Astronomy, Stockholm University, AlbaNova University Center, SE-106 91 Stockholm, Sweden}
\altaffiltext{10}{Department of Physics and Astronomy, College of Charleston, 66 George Street, Charleston, SC}
\altaffiltext{11}{Department of Mechanical and Aerospace Engineering, Princeton University, Princeton, NJ}
\altaffiltext{12}{Exoplanets and Stellar Astrophysics Laboratory, Code 667, NASA-Goddard Space Flight Center, Greenbelt, MD}
\altaffiltext{13}{Department of Astronomy, The University of Tokyo, 7-3-1 Hongo, Bunkyo-ku, Tokyo, Japan }
\altaffiltext{14}{Chile Observatory, National Astronomical Observatory of Japan, Osawa, Mitaka, Tokyo, Japan}
\altaffiltext{15}{Eureka Scientific, 2452 Delmer, Suite 100, Oakland, CA}
\altaffiltext{16}{Department of Astrophysical Sciences, Princeton University, Princeton, NJ}
\altaffiltext{17}{Institute of Space and Astronautical Science, JAXA, 3-1-1 Yoshinodai, Sagamihara, Kanagawa, Japan}
\altaffiltext{18}{National Meteorological Satellite Center, Jincheon, Chungbuk, 27803, Republic of Korea}
\altaffiltext{19}{Homer L. Dodge Department of Physics, University of Oklahoma, Norman, OK}
\altaffiltext{20}{Department of Physics, University of Cincinnati, Cincinnati, OH}
\begin{abstract}
We present $H$-band scattered light imaging of a bright debris disk around the A0 star HD 36546 obtained from the Subaru Coronagraphic Extreme Adaptive Optics (SCExAO) system with data recorded by the HiCIAO camera using the vector vortex coronagraph.  SCExAO traces the disk from $r$ $\sim$ 0\farcs{}3 to $r$ $\sim$ 1\arcsec{} (34--114 au).  The disk is oriented in a near east-west direction (PA $\sim$ 75${\arcdeg}$), is inclined by $i$ $\sim$ 70--75${\arcdeg}$ and is strongly forward-scattering ($g$ $>$ 0.5).   It is an extended disk rather than a sharp ring; a second, diffuse dust population extends from the disk's eastern side.    While HD 36546 intrinsic properties are consistent with a wide age range ($t$ $\sim$ 1--250 $Myr$), its kinematics and analysis of coeval stars suggest a young age (3--10 $Myr$) and a possible connection to Taurus-Auriga's star formation history.    SCExAO's planet-to-star contrast ratios are comparable to the first-light \textit{Gemini Planet Imager} contrasts; for an age of 10 $Myr$, we rule out planets with masses comparable to HR 8799 b beyond a projected separation of 23 au.  A massive icy planetesimal disk or an unseen superjovian planet at $r$ $>$  20 au may explain the disk's visibility.
The HD 36546 debris disk may be the youngest debris disk yet imaged, is the first newly-identified object from the now-operational SCExAO extreme AO system, is ideally suited for spectroscopic follow up with SCExAO/CHARIS in 2017, and may be a key probe of icy planet formation and planet-disk interactions.
\end{abstract}
\section{Introduction}
Cold debris disks around nearby, young stars offer a reference point for the formation and evolution of the Kuiper belt and provide evidence for unseen planets \citep{Wyatt2008}.   
Debris disk luminosities are highest at the youngest ages (5--30 $Myr$) around stars more massive than the Sun; the luminosity of these debris disks  
may trace debris production from collisions between boulder-sized planetsimals as a byproduct of icy planet formation \citep[``self-stirring",][]{KenyonBromley2008,Currie2008}.
Unseen massive planets may also dynamically stir icy planetesimals to make debris disks visible and sculpt debris disks \citep[``planet stirring",][]{Mustill2009}.  

Resolved images of debris disks probe icy planet formation and reveal evidence for hidden planets \citep{Schneider2009,Currie2015a}.  In some cases, planets stirring debris disks were subsequently imaged; the properties of the debris disks help constrain the masses of planets \citep[e.g.][]{Lagrange2010,Rodigas2014,Nesvold2015}.    As nearly all of these resolved debris disks surround stars older than $\sim$ 10 $Myr$ and most protoplanetary disks dissipate by $\sim$ 3--5 $Myr$ \citep{Cloutier2014,Choquet2015}, resolved images of debris disks around stars younger than 10 $Myr$ shed new light on icy planet formation and planet-debris disk interactions for the youngest, fully-formed planetary systems.


HD 36546 is a B8--A0 star located slightly foreground  \citep[$d$ = 114 pc,][]{vanLeewen2007} to the 1--2 $Myr$ old Taurus-Auriga star-forming region \citep[$d$ $\sim$ 140 pc,][]{Kenyon2008,Luhman2009} and a promising new target around which to search for young exoplanets and planet-forming disks.   
The star has extremely strong mid-to-far infrared excesses -- among the largest of newly-identified WISE debris disk candidates studied in \citet{Wu2013} -- suggestive of copious circumstellar dust.   
 Its fractional disk luminosity ($L_{\rm IR}$/$L_{\star}$ $\sim$ 4$\times$10$^{-3}$) rivals that of benchmark resolved debris disk-bearing systems such as $\beta$ Pictoris, HR 4796A, and HD 115600 \citep{SmithTerrile1984,Schneider2009,Currie2015a}.

In this Letter, we report spatially-resolved imaging of HD 36546's debris disk from the Subaru Coronagraphic Extreme Adaptive Optics system \citep{Jovanovic2015} on the 8.2 m Subaru Telescope on Maunakea.   
The HD 36546 debris disk is the first newly-identified object from the now-operational SCExAO extreme AO system and potentially the youngest debris disk ever spatially resolved in scattered light.


\section{SCExAO Observations and Data Reduction}
Given its extremely large infrared excess, HD 36546 had long been (since 2013) a prime direct imaging target for SCExAO once extreme AO capability had been achieved.  
 Following a successful July 2016 engineering run where SCExAO achieved $H$-band Strehl ratios of $\sim$ 80\% on sky  \citep{Jovanovic2016},
 we targeted the star during the following run, on 15 October 2016, also in $H$ band using the HiCIAO infrared camera and the vector vortex coronagraph (Kuhn et al. in prep.) and in angular differential imaging mode \citep{Marois2006}.   SCExAO ran at 2 kHz,  correcting for 1080 modes.  Despite ``fast", poor (for Maunakea) atmospheric conditions ($\theta$ $\sim$ 1.0\arcsec{} seeing, 12 m/s wind), skies were clear and SCExAO successfully closed loop, yielding $H$-band Strehl ratios of 70--80\% on HD 36546 and digging out a dark hole in the stellar halo interior to $r$ $\sim$ 0\farcs{}8.

HD 36546 exposures consisted of co-added 30 $s$ frames where the detector response was linear exterior to $r$ $\sim$ 0\farcs{}1; the observations totaled 42 minutes of integration time and were centered on transit, yielding 113${\arcdeg}$ of parallactic motion (4.7 $\lambda$/D at 0\farcs{}1).   For photometric calibration we obtained unsaturated exposures of HD 48097 using the neutral density filter just prior to HD 36546.   For astrometric calibration (distortion, north position angle), we observed the M15 globular cluster.   The distortion-corrected images have a pixel scale of 8.3 mas pixel$^{-1}$.
Basic image processing steps followed those employed by \citet{Garcia2016} for SCExAO/HiCIAO data, including de-striping, bad pixel masking/correction, flat fielding, distortion correction, and precise (to fractions of a pixel) image registration.   

We performed point-spread function (PSF) subtraction using the A-LOCI pipeline \citep{Currie2012}, which builds upon the original locally-optimized combination of images (LOCI) algorithm \citep{Lafreniere2007a}, and utilizes a moving pixel mask to reduce the signal loss induced by the algorithm and a singular value decomposition (SVD) cutoff to reduce errors propagating through the matrix inversion \citep[][]{Marois2010b,Currie2012}.    
To optimize our ability to detect disks, we altered the geometry of the subtraction zone (region of the image to subtract at a given time) and optimization zone (region from which reference image coefficients used to build up a reference PSF are determined).  We defined the optimization zone as a ring of width 10 pixels and the subtraction zone as a wedge-like section of this ring, a setup found to sometimes yield better detections of edge-on disks.

\section{Detection and Basic Morphology of the HD 36546 Debris Disk}

\begin{figure}
\centering
\includegraphics[scale=0.33,trim=0mm 0mm 0mm 0mm,clip]{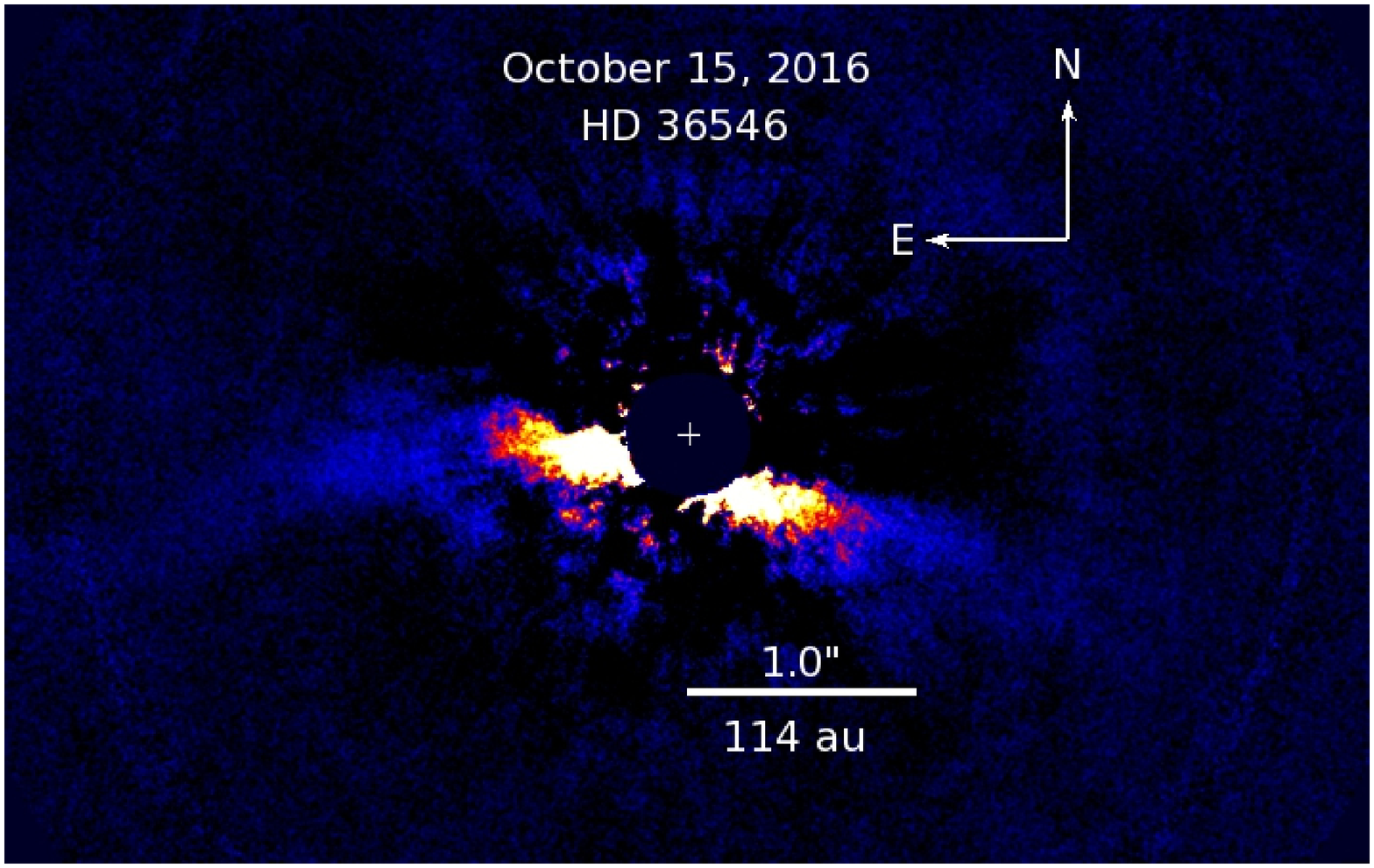}
\includegraphics[scale=0.33,trim=0mm 0mm 0mm 0mm,clip]{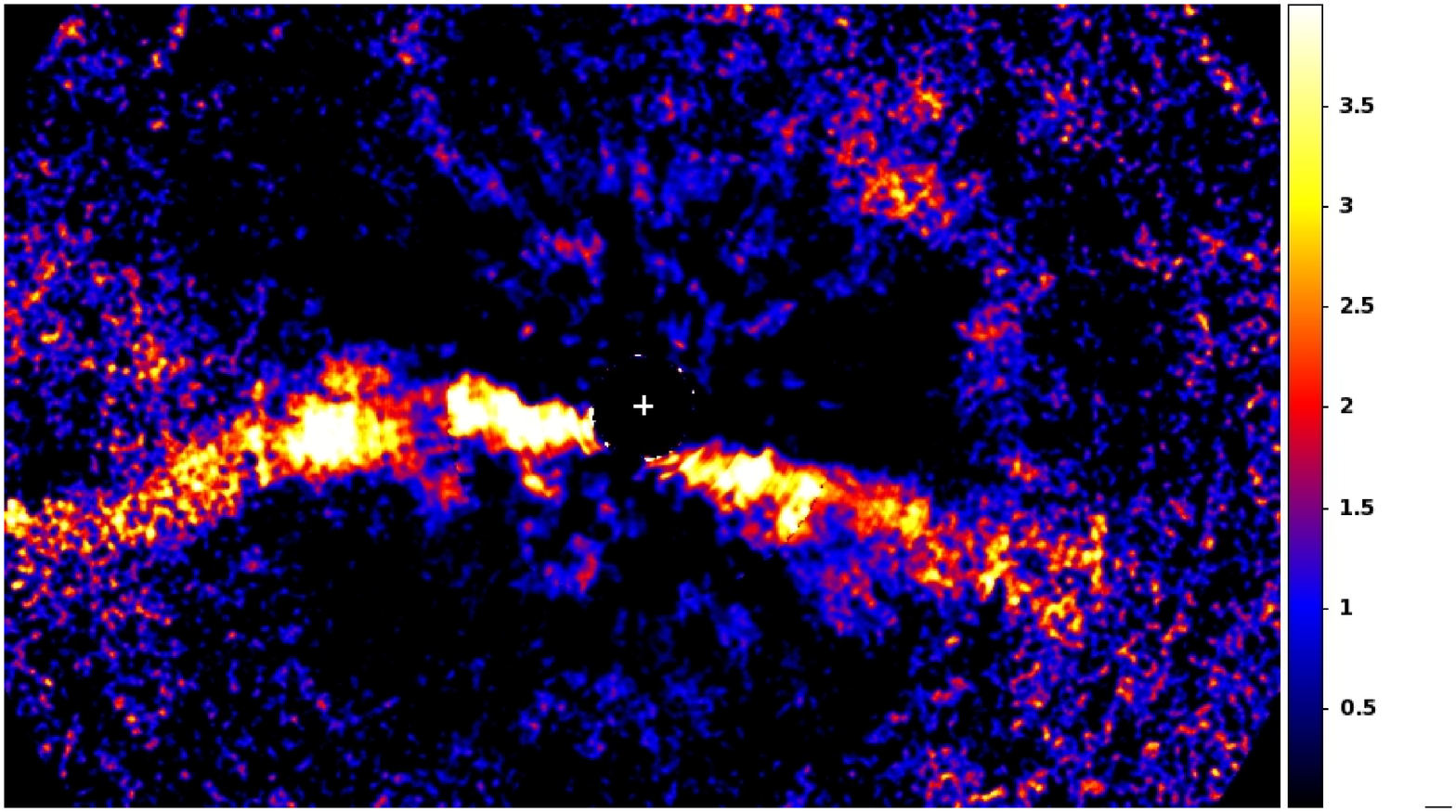}
\caption{(left) Detection of the HD 36546 debris disk with SCExAO/HiCIAO.  The inner $r$ $\le$ 0\farcs{}3 from the star's position (cross) is masked.  We imposed a rotation gap of $\delta$=0.8$\times$FWHM, used an SVD cutoff of 10$^{-4}$, and utilized all available reference images.  A wide range of settings resulted in a statistically significant detection (SNRE $\gtrsim$ 3 along the disk spine): e.g. $\delta$ = 0.4--2.5, SVD$_{\rm lim}$ = 10$^{-1}$--10$^{-7}$.     (right) Signal-to-noise map showing that the detection of HD 36546's disk is statistically significant.
}
\label{images}
\end{figure}

Figure \ref{images} (left panel) displays the combined, PSF-subtracted image (linear stretch) plainly revealing a debris disk around HD 36546  with a near-east/west orientation, extending from 0\farcs{}3 to 1\arcsec{} ($r$ $\sim$ 34--114 au) and diffuse emission extending from the east disk ansae and visible above the background out to 3\arcsec{}.    The trace of the disk is offset from the star's position, suggesting that the disk is not viewed perfectly edge on and/or is strongly forward-scattering, similar to some well-studied debris disks like HD 32297 \citep[e.g.][]{Rodigas2014b}.   

To estimate the disk's signal-to-noise per resolution element (SNRE), we followed the standard approach \citep{Currie2011a} of replacing each pixel with the sum of values enclosed by a FWHM-wide aperture ($r_{\rm ap}$ $\sim$ 2.5 pixels) but masked the visible trace of the disk when computing the noise at a given angular separation.  The spine of the main disk is over 3--5 $\sigma$ significant on both sides from 0\farcs{}3 to 1\farcs{}1 (Figure \ref{images}, right panel), peaking at 8-$\sigma$\footnote{We likely detect disk signal down to $r$ $\sim$ 0\farcs{}15 (not shown), but the SNRE ($\sim$ 2-3) is too low to be decisive.}.   

\section{Analysis}
\subsection{Disk Geometry}
To determine the disk's position angle, we followed analysis employed for the $\beta$ Pic debris disk in \citet{Lagrange2012}, using ``maximum spine" and Lorentzian profile fitting.  We performed fits using the IDL \textrm{mpfitellipse} package, where the pixels are weighted by their (``conservatively" estimated) SNRE, focusing on regions where the disk is detected at SNRE $\ge$ 3 according to our conservative estimate of SNRE and at separations of $r$ = 0\farcs{}3--1\farcs{}0.  
Lorentzian profile fitting yields a position angle of 74.4$^{o}$ $\pm$ 0.8$^{o}$.   ``Maximum spine" fitting yields nearly identical results: 75.3${\arcdeg}$ $\pm$ 0.5${\arcdeg}$.  

\subsection{Disk Forward Modeling}
Following the same analysis performed for HD 115600 \citep{Currie2015a}, 
we inferred additional disk properties by generating a grid of synthetic scattered light images produced using the GRaTeR code \citep{Augereau1999}.  
After convolving the model disk with the mean unsaturated PSF constructed from our photometric standard, we inserted each model disk into a sequence of empty images with position angles identical to those from the HD 36546 observations.  
  We then performed PSF subtraction on synthetic images containing the model disk using the same A-LOCI coefficients that were applied to the real data and compared the attenuated, synthetic disk image with the real disk image.  

Table \ref{diskmodels} (left two columns) describes our model parameter space.  We adopted the position angle determined above (75$^\circ$) and assumed a zero eccentricity for simplicity but varied other parameters.   The visible trace of the disk drops off by $r$ $\sim$ 0\farcs{}65 ($\sim$ 74 au) in projected separation, substantially exterior to 0\farcs{}85 ($\sim$ 97 au); the model parameter space covers disk stellocentric distances of $r_{o}$ = 75--95 au.  From inspection, the disk scale height at these locations ($r_{o}$) has to be at least 5 au but unlikely to be more than 15 au to be consistent with the self-subtracted images.  Thus, we considered heights of $ksi_{\rm o}$ = 5--15 au.   We varied the Henyey-Greenstein scattering parameter $g$ (0--0.85) and density power laws describing the decay of disk emission away from the photocenter ($\alpha_{\rm in}$ = 3, 5, 10; $\alpha_{\rm out}$ = $-1$, $-3$, $-$5, $-$10).
Values outside our adopted ranges (e.g. $ksi_{\rm o}$ = 1 au, $i$ = 65${\arcdeg}$) yielded processed synthetic disk images strongly discrepant with the real data and are not considered.
  
 Following \citet{Thalmann2013}, we assessed the fidelity of the model disk to the observed disk by comparing the residuals of images subtracted by the model disk and binned by 1 FWHM.   We defined our ``region of interest" from which we quantify the residuals by the visible trace of the disk 
  between 0\farcs{}3 and 1\farcs{}0.
``Acceptably fitting" models fulfill $\chi^{2}$ $\le$ $\chi^{2}_{\rm min}$ + $\sqrt{2\times N_{\rm data}}$ \citep{Thalmann2013}, where N$_{\rm data}$ is 505 binned pixels.   
The $\chi^{2}_{\nu}$ threshold for identifying acceptably-fitting models is $\chi^{2}_{\nu}$ $\sim$ 1.065.

\begin{figure}
\centering
\includegraphics[scale=0.2,trim=0mm 0mm 20mm 0mm,clip]{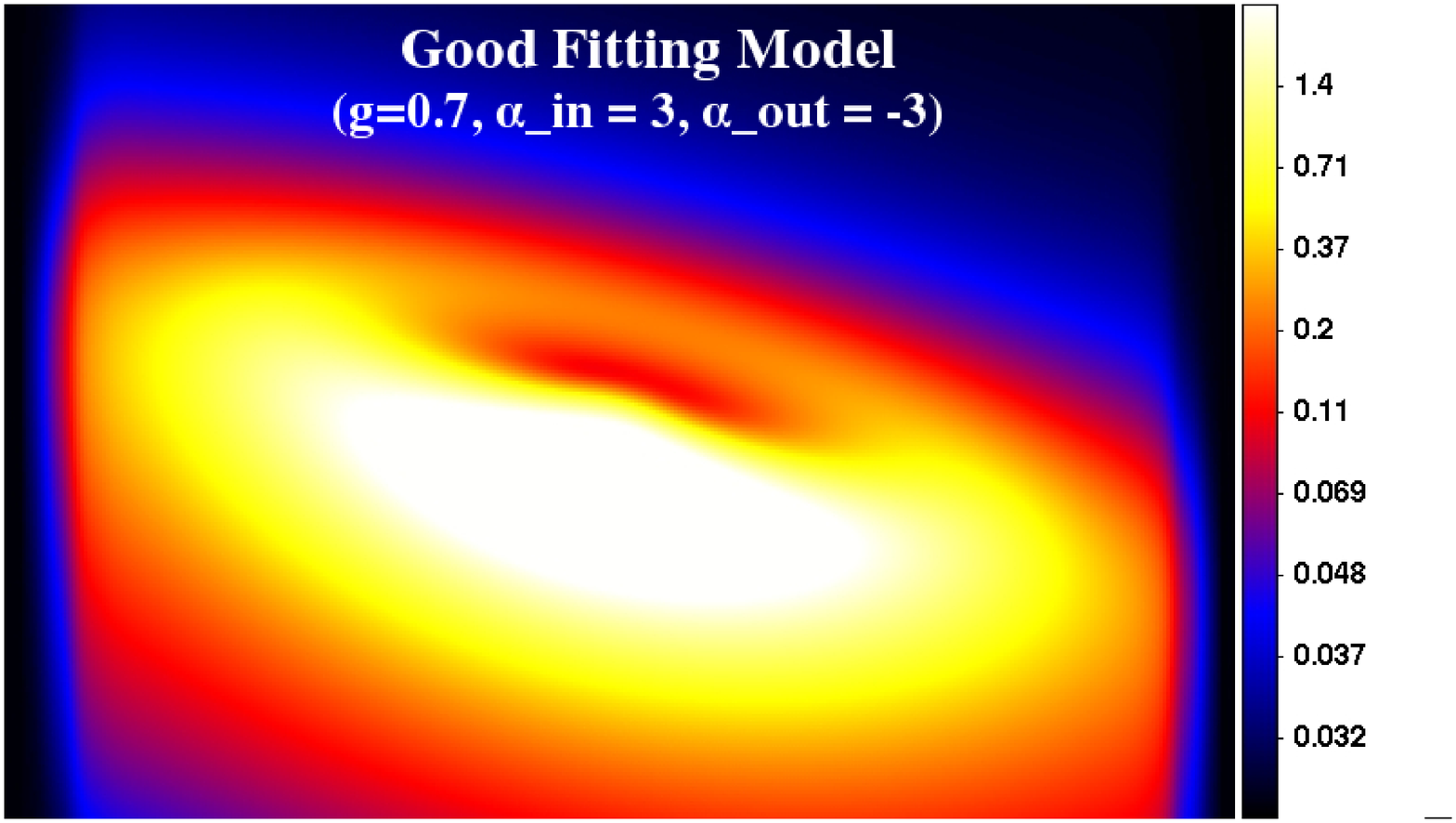}
\includegraphics[scale=0.2,trim=0mm 0mm 20mm 0mm,clip]{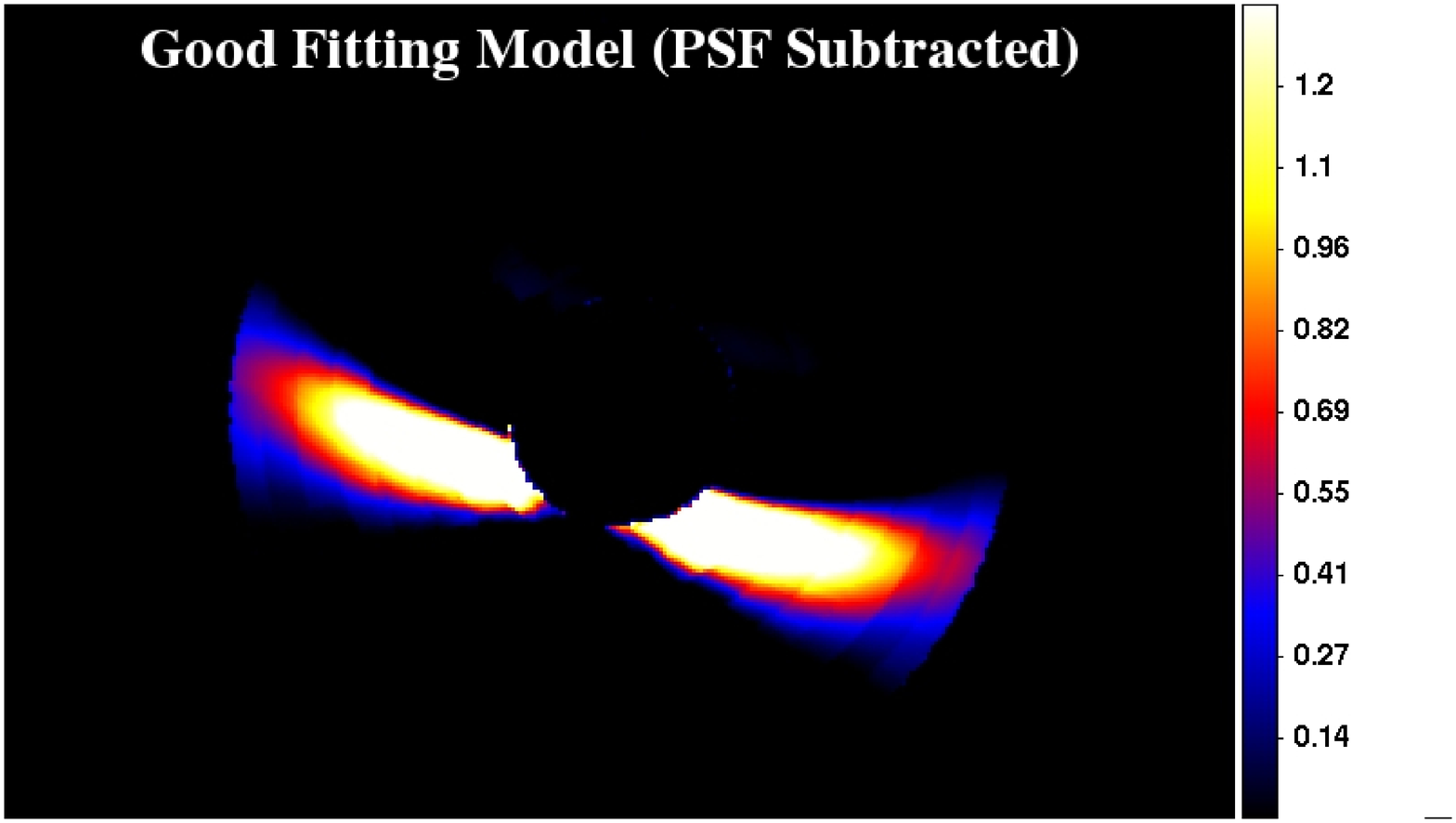}
\includegraphics[scale=0.2,trim=0mm 0mm 10mm 0mm,clip]{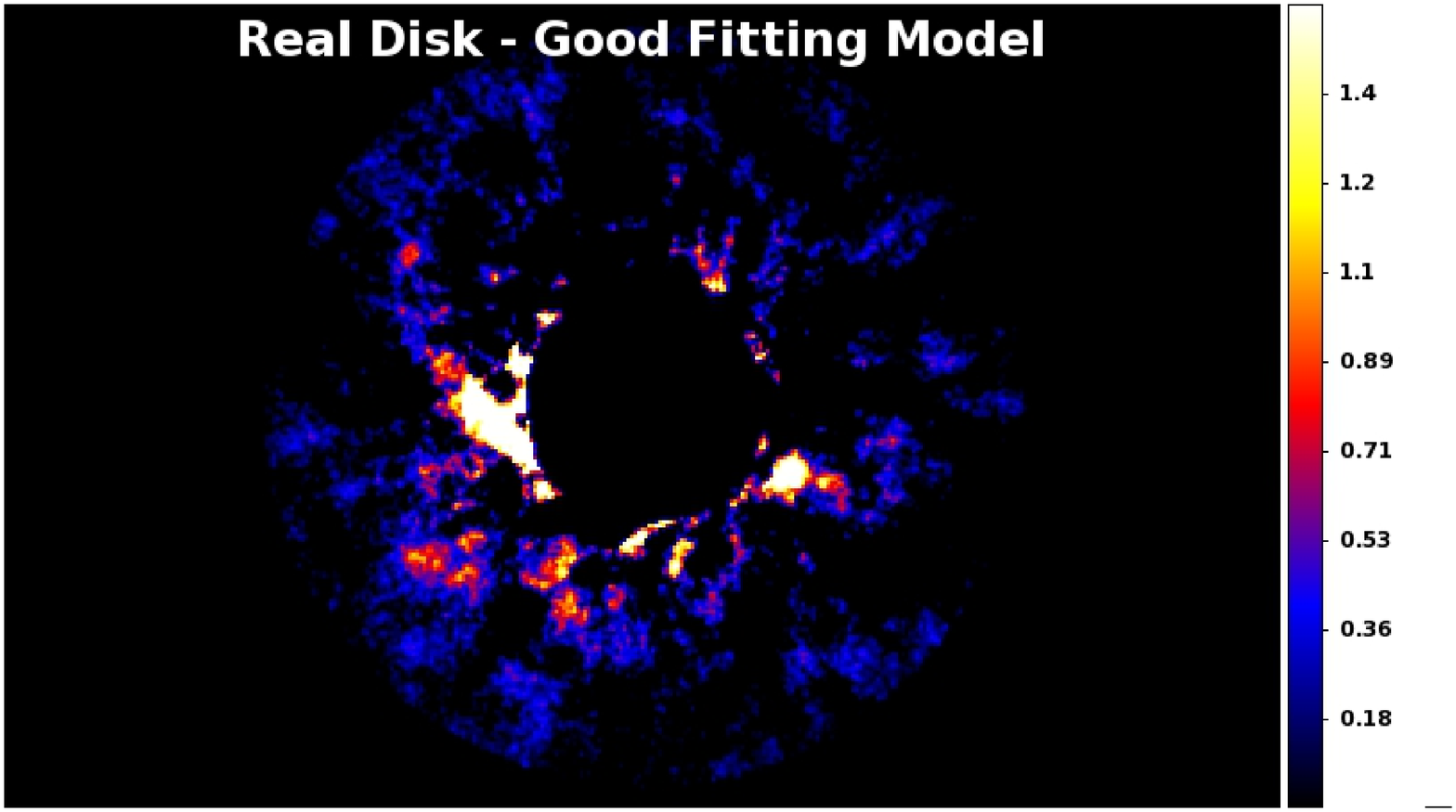}
\caption{Forward-modeling of HD 36546's disk emission.   All panels are units of $mJy$/$arcsec^{2}$ (see vertical color bars).  The left panel shows an input acceptably-fitting model -- $g$ = 0.7, $ksi_{\rm o}$ = 5 au, $r_{\rm o}$ = 95 au, $i$ = 75${\arcdeg}$, $\alpha_{\rm in}$ = -3, and $\alpha_{\rm out}$ = 3 ($\chi^{2}_{\nu}$ = 1.03), the middle panel shows the simulated PSF-subtracted model, and the right panel shows the residuals of the real minus simulated model subtraction.   The residuals at small angular separation reveal a slight mismatch in reproducing the disk's self-subtraction footprints at separations where the disk's SNRE is low. 
  }
\label{model_images}
\end{figure}

Table 1 (third and fourth columns) describes our modeling results and Figure \ref{model_images} displays one of our acceptably-fitting models.   The best-fitting model is a strongly forward-scattering ($g$ = 0.85) disk inclined by $i$ = 75${\arcdeg}$ centered on 85 au with a FWHM ($ksi_{\rm o}$) of 10 au and modest power-law decays in its density (abs($\alpha_{\rm in,out}$) = 3).   The family of acceptably fitting models exclusively draw from forward-scattering disks ($g$ $\sim$ 0.7--0.85) inclined by 70--75${\arcdeg}$ with a shallow power law decay at large distances ($\alpha_{\rm out}$ = -3), suggestive of an extended disk rather than a narrow ring.   
In contrast, disk models with scattering properties and morphologies comparable to well-known debris disks HR 4796A and HD 115600\citep{Schneider2009,Currie2015a} are inconsistent with the HD 36546 disk image.   Simulated PSF-subtracted disk models with $g$ $\le$ 0.5 and/or sharp outer disk power laws ($\alpha_{\rm out}$ $\le$ 5) incorrectly predict that both sides of the disk and/or the disk ansae are detectable.  

\subsection{Analysis of HD 36546: Spectral Type, Age, and Membership}
To better understand HD 36546 within the general context of planet formation, we (re-)assessed the primary star's spectral type, age, and evidence for membership to known moving groups/star formation events.  

While some authors \citep[e.g.][]{Chini2012} list HD 36546 as a B8 star, others claim the star has an A0 type \citep[e.g.][]{Abt2004}.  
We independently spectral typed HD 36546 from publicly-available, processed \textit{FAST} archive spectra\footnote{\url{http://tdc-www.harvard.edu/cgi-bin/arc/fsearch}} taken on 2007 February 9 (Program 164) using the SPTCLASS code \citep{Hernandez2004}.

\begin{figure}
\centering
\includegraphics[scale=1,trim=0mm 2mm 51mm 2mm,clip]{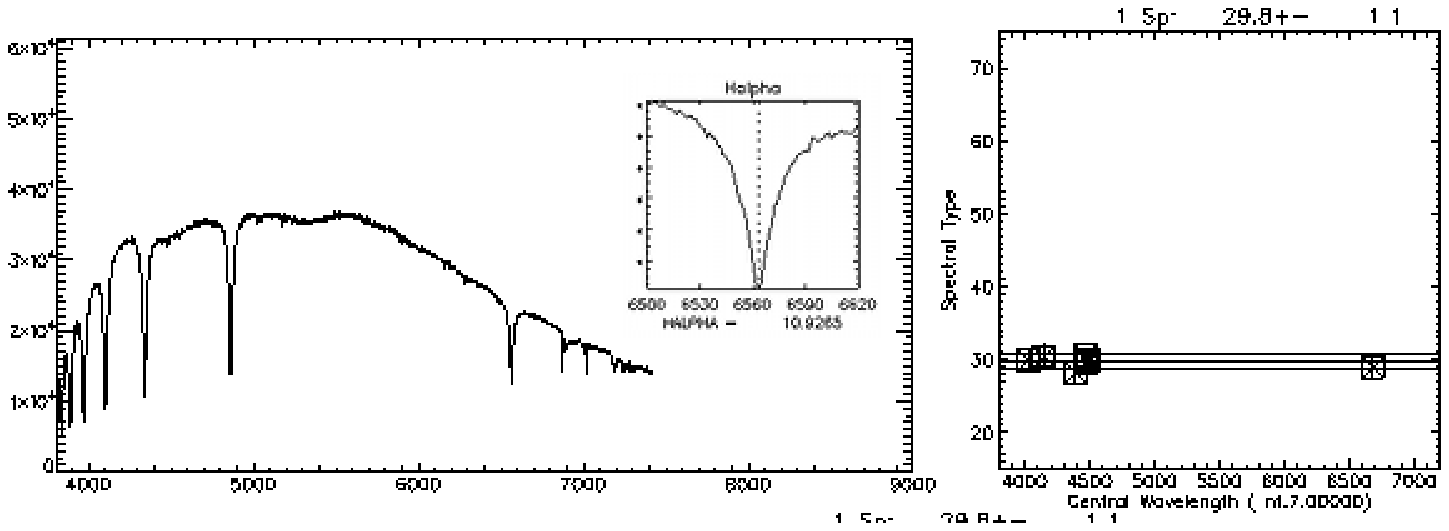}
\includegraphics[scale=0.23,trim=0mm 6mm 0mm 6mm,clip]{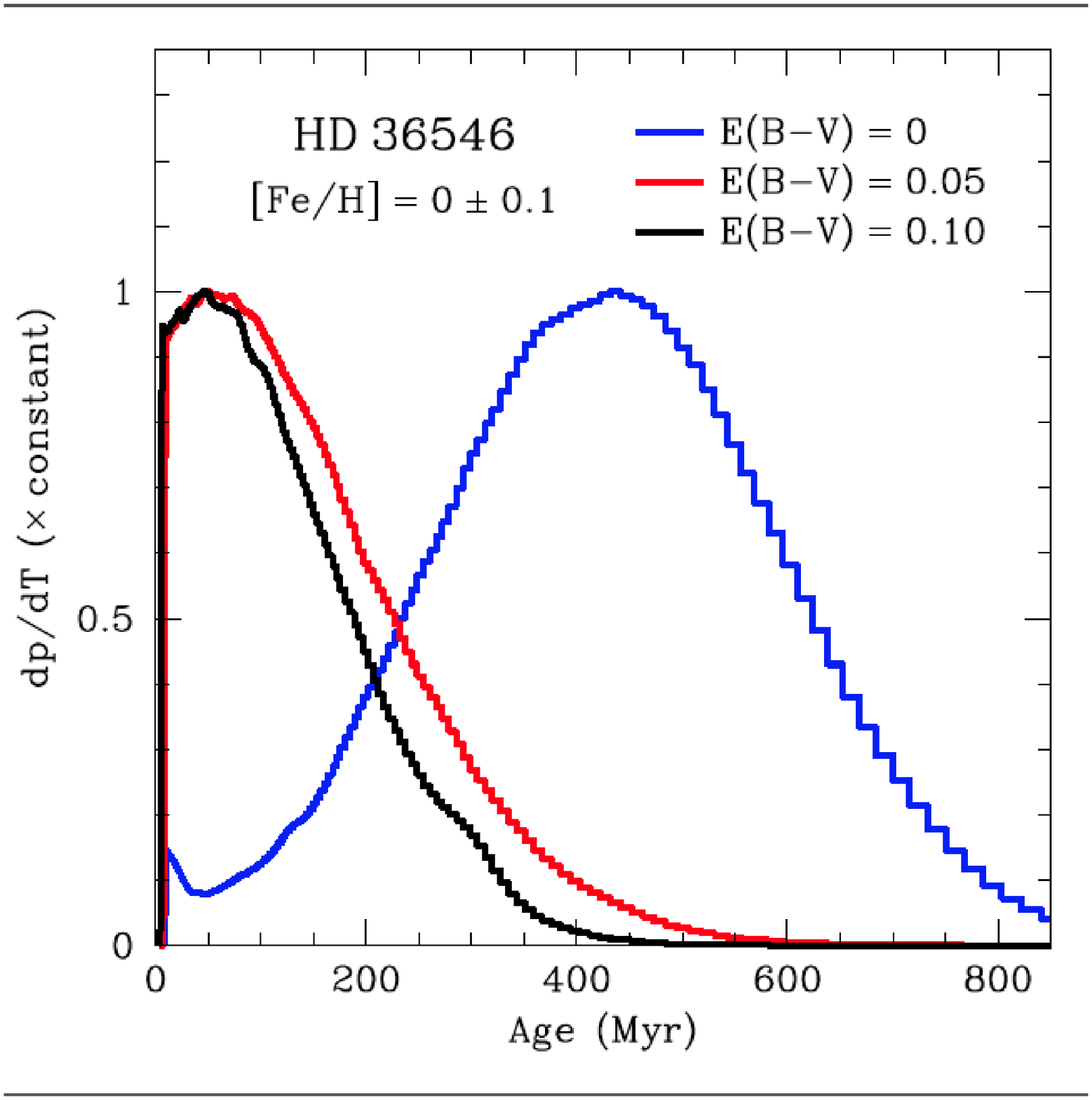}
\caption{Analysis of HD 36546's spectral type and age.  (left) The star's FAST archive spectrum.   The inset shows the $H\alpha$ line and equivalent width in angstroms (10.9 \AA), consistent with being an A0 star (EW($H\alpha$) $\sim$ 10--11 \AA\ for an A0 star vs $7-9$ \AA\ for a B7--B9 star\footnote{\url{http://dept.astro.lsa.umich.edu/$\sim$hernandj/SPTclass/H$\_$alpha.ps.gif}}).   Other Balmer and He I line strengths favor an A0 spectral type.  (right) The \citep{BrandtHuang2015} Bayesian analysis showing that HD 36546's age is 1--250 $Myr$ old given its reddening (E(B-V) = 0.06), consistent with a 3--10 $Myr$ age estimated from membership in Mamajek 17.}
\label{ageest}
\end{figure}

 HD 36546 is a textbook A0V star whose Balmer (He) lines are too strong (weak) to be a B8 star (Figure \ref{ageest}, left panel).   The star exhibits no $H\alpha$ emission line reversal suggestive of gas accretion, consistent with its lack of warm excess (as probed by WISE [3.4]-[4.6] colors).   HD 36546A's Tycho-II catalog photometry ($B-V$ = 0.07; \citealt{Hog2000}) and intrinsic A0V star colors \citep{Pecaut2013} imply a reddening of $E(B-V)$ $\sim$ 0.06.

Direct age estimates for HD 36546, although poorly constrained, are broadly consistent with a young age expected for a star surrounded by an extremely dusty disk.    For a reddening comparable to our derived value, the \citet{BrandtHuang2015} Bayesian method implies a possible age range of $t$ $\sim$ 1--250 $Myr$ (68\% confidence interval).

HD 36546 lies foreground (by 14 pc) to the Taurus-Auriga molecular cloud \citep[$d$ = 127-147 pc][]{Torres2007}.  Recent analysis suggests that the star is a member of an association labeled \textit{Mamajek 17}
 \citep{Mamajek2016}, which also includes 
 weak-line T Tauri stars identified originally by \citet{Li98} and 
 mid-M stars studied by \citet{Slesnick2006a} (J0539009+2322081, J0537385+2428518).   
Mamajek 17 may be an earlier epoch of star formation within the wider Taurus-Auriga complex.

  Based on the strengths of the gravity-sensitive Na I 8190 $\dot{A}$ line, \citeauthor{Slesnick2006a} conclude that these M stars are younger than Upper Scorpius \citep[age $\approx$ 10 $Myr$,][]{Pecaut2012} and comparable in age to Taurus-Auriga (1--2 $Myr$).   Inspection of Figure 4 in \citet{Slesnick2006a} and Figure 11 in \citet{Slesnick2006b}, though, implies that some mid-M stars have Na I strengths comparable to Upper Scorpius stars.   Furthermore, the mid-M star members studied by \citeauthor{Slesnick2006a} have measured $J$-$K$ colors implying little reddening and a lack of associated diffuse material expected for 1--2 $Myr$-old stars.
On the other hand, lithium lines in Mamajek 17's mid-K stars have equivalent widths of $\approx$ 500 m$\dot{A}$, significantly larger than those for 30--120 $Myr$-old mid-K stars but similar to stars in the 3--8 $Myr$ old $\epsilon$ and $\eta$ Cha associations \citet[][]{Murphy2013}.   

To independently assess HD 36546's possible membership in this group, we computed and compared HD 36546's UWV space motions to the association's mean value (minus HD 36546).  Provided that their distances and radial-velocities are similar (within $\sim$ 5--10 $pc$ and 2 km s$^{-1}$), HD 36546 and the other Mamajek 17 members share indistinguishable space motions: $U,V,W$ $\sim$ 18.9 $\pm$ 2.0, 22.4 $\pm$ 0.1, and 9.6 $\pm$ 0.2 km s$^{-1}$ for the star and 16.1 $\pm$ 1.0, 21.5 $\pm$ 2.4, and 11.2 $\pm$ 2.2 km s$^{-1}$ for the others.
    
Our simple analysis then suggests that HD 36546 is a likely member of the proposed Mamajek 17 group and thus its age is likely 3--10 $Myr$\footnote{A more detailed analysis of
   Mamajek 17's low-mass stars favor an age of 6 Myr over older and (especially) younger ages found to be acceptable for member stars in this work (E. Mamajek 2016, pvt. comm.).}.
  
  \subsection{Planet Detection Limits and the Source of HD 36546's Debris Disk}
  To place limits on jovian planets plausibly responsible for stirring the HD 36546 debris disk, we reprocess our image sequence using a (different) set of algorithm parameters that maximize the SNR of point sources.   To determine this optimal parameter space and then derive a contrast curve, we iteratively input, simulate the subtraction, and measure the output SNR of point sources over our angular separation of interest ($r$ $\sim$ 0.15--0\farcs{}9: $\sim$ 3 $\lambda$/D to the plausible disk inner edge) along the major disk axis.   
  %

\begin{figure}
\centering
\includegraphics[scale=0.5,clip]{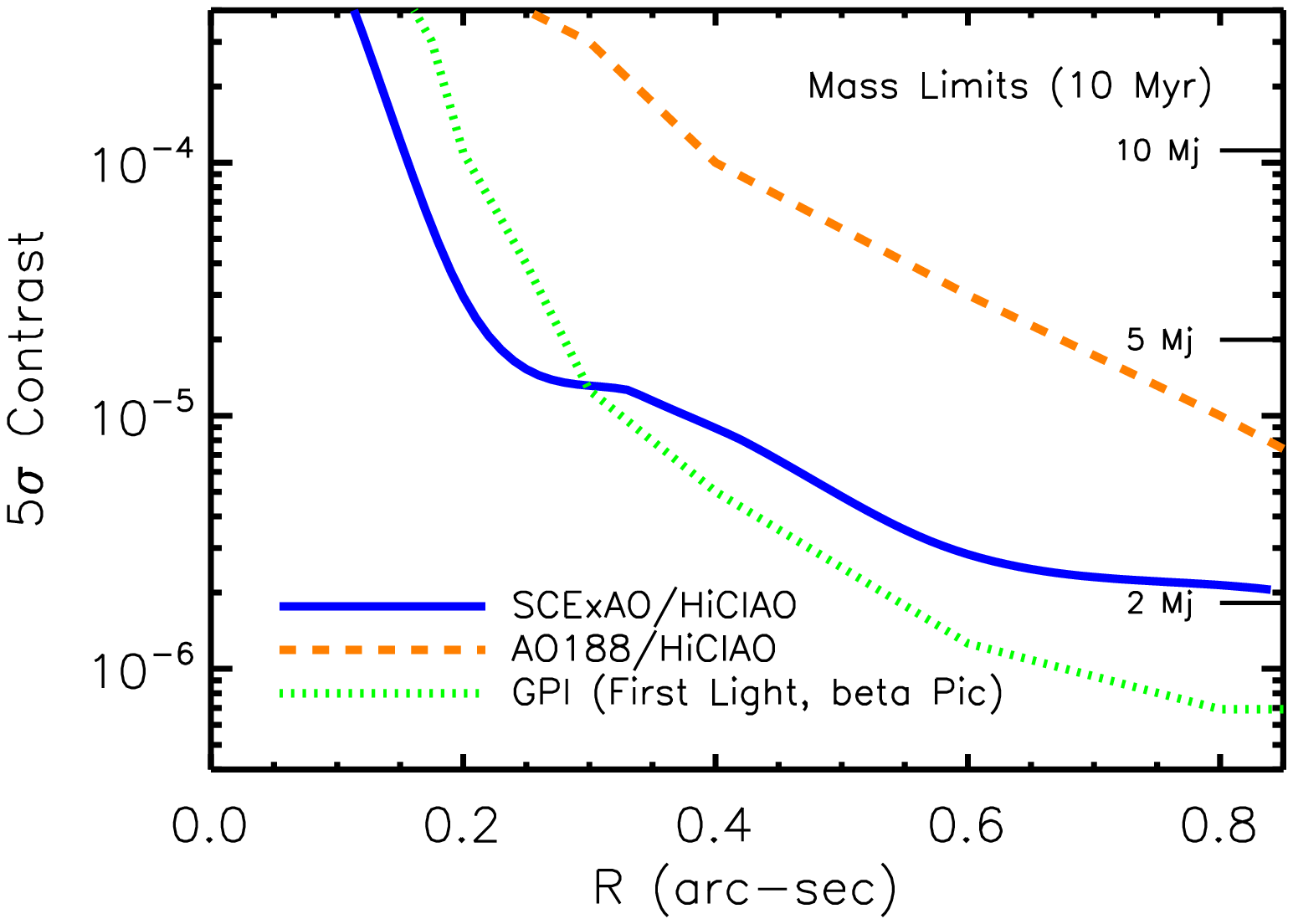}
\includegraphics[scale=0.5,clip]{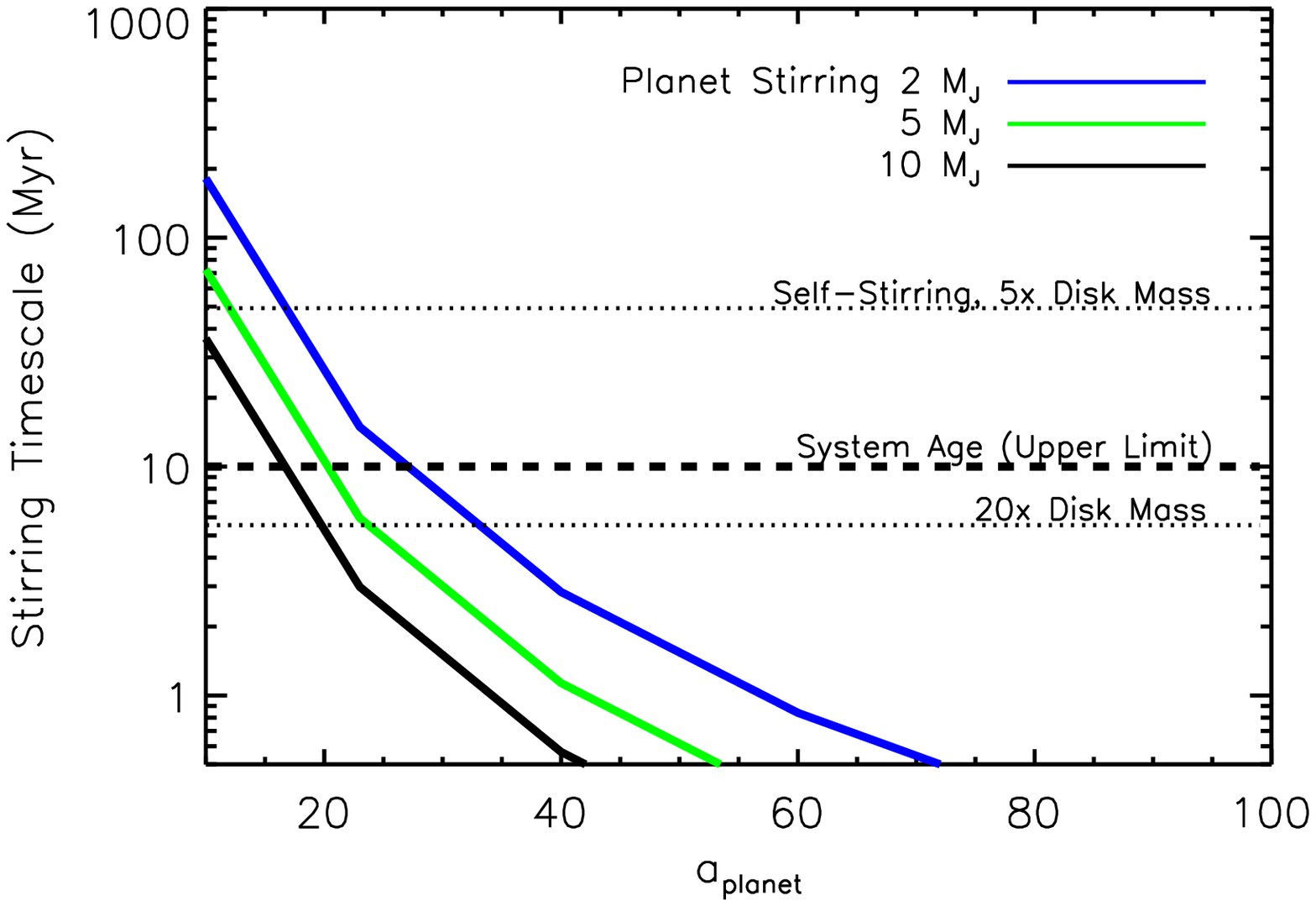}
\caption{ (left)
SCExAO contrast limits for HD 36546 compared to the best contrasts from Subaru/HiCIAO using the AO188 facility AO system (orange, from \citealt{Brandt2014}), first-light contrasts from the \textit{Gemini Planet Imager} \citep{Macintosh2014}, and predicted contrasts for 10 $Myr$-old planets between 2 and 10 $M_{J}$ from the \citet{Baraffe2003} luminosity evolution models (horizontal bars).   We adopted a rotation gap of $\delta$ = 0.44 and an SVD cutoff of 1$\times$10$^{-6}$, while selecting only the 66 best-correlated images (within each optimization area) for PSF subtraction.     (right) Mechanisms for stirring HD 36546's debris ring.    Only for an exceptionally massive disk does self-stirring (horizontal dotted lines) as modeled in \citet{KenyonBromley2008} occur in less than 10 $Myr$ (the system's age if in Mamajek 17).  Planet stirring (solid lines; Mustill \& Wyatt 2009) assuming an planet eccentricity of $e$ = 0.1 can explain the debris disk if the planet orbits at $r$ $\gtrsim$ 20 au.  Assuming $e$ = 0.01, the planet would have to orbit beyond 35 au (not shown).}
.
 
\label{masslimit}
\end{figure}

Figure \ref{masslimit} (left) shows the resulting contrast curve and planet mass sensitivity limit (corrected for finite sample sizes as in \citealt{Mawet2014}).  Despite poor observing conditions, SCExAO/HiCIAO achieves 5-$\sigma$ planet-to-star contrasts of 2.2$\times$10$^{-6}$,  8.3$\times$10$^{-6}$, 3$\times$10$^{-5}$, and 1.7$\times$10$^{-4}$ at $r$ $\sim$ 0\farcs{}75, 0\farcs{}4,  0\farcs{}2, and 0\farcs{}14.    At small ($r$ $\lesssim$ 0\farcs{}3) separations, SCExAO's performance is comparable to the first-light performance of the \textit{Gemini Planet Imager} \citep{Macintosh2014}.   Compared to conventional AO imaging with Subaru, SCExAO yields a factor of 10--20 contrast gain at $r$$\sim$ 0\farcs{}2--0\farcs{}6.
 
 Assuming an age of 10 $Myr$, our data rules out the presence of planets with masses greater than 5--6 $M_{J}$, comparable to HR 8799 b \citep{Marois2008,Currie2011a}, at projected separations of 23 au ($r$ $\sim$ 0\farcs{}2) and 2.5 $M_{J}$ planets at wide ($r$ $\gtrsim$ 0\farcs{}6 or 70 au) projected separations.  For an age of 250 $Myr$, our data still rule out analogues to ROXs 42Bb ($M$ $\sim$ 9 $M_{J}$; \citealt{Currie2014}) beyond $r$ $\sim$ 70 au (not shown).  
 
 To assess whether HD 36546's disk can be explained by ``planet stirring" or ``self-stirring" of icy planetesimals, we followed steps in \citet{Mustill2009}, comparing the stirring timescale for planets of different masses and timescales for icy planetesimal disks of different masses assuming an 85 au disk radius (Figure \ref{masslimit}, right)\footnote{Since HD 36546's disk is an extended disk, we do not utilize planet mass/location estimates from the morphology of ring-like disks as in \citet{Rodigas2014}.}.   For a system age of 10 $Myr$, self-stirring requires a planetesimal disk 15--20 times more massive than the nominal value adopted in \citet{KenyonBromley2008}.   
However, a 2--10 $M_{\rm J}$ planet with an eccentricity of $e$ = 0.1 orbiting beyond 20 au could explain the disk.   While we fail to detect such a planet, it could be positioned along the disk's minor axis and thus at smaller projected separations where SCExAO's sensitivity is poorer.
\section{Discussion}

   While many early-type stars younger than $\approx$ 8 $Myr$ show evidence for a debris disk, until now arguably none have been resolved \citep[][]{Currie2008, Choquet2015}\footnote{The nature of the circumstellar environment for 5--7.5 $Myr$-old HD 141569A is unclear, as its inner regions resemble that of a transitional protoplanetary disk \citep[see][]{Currie2016}.}.  
   At 3--10 $Myr$ (assuming membership in Mamajek 17), HD 36546's debris disk could be the youngest resolved debris disk to date.
   
  Measurements of HD 36546's rotation rate and inclination could independently the star's age and thus solidify the interpretation of the system within the context of planet formation.   Ages derived for early-type stars based on HR diagram measurements are extremely sensitive to rotation/inclination effects.   As shown for $\kappa$ And, including rotation/inclination effects can significantly reduce the star's estimated age and the interpretation of any imaged companions
  \citep{Carson2013,Jones2016}.

  The brightness of HD 36546's debris disk and its extremely advantageous (for observing from Maunakea) declination, make it an obvious target for spectroscopic follow-up with the CHARIS integral field spectrograph \citep{Groff2015}.   While even single-band (e.g. $H$ band) disk spectra shed some light on debris disk grain compositions \citep[][]{Currie2015a}, 
   simultaneous $JHK$ spectra available with CHARIS will allow more decisive constraints.    
  
  Within the next year, SCExAO should be consistently achieving Strehl ratios of $\sim$ 90\%  and, with CHARIS and advanced image processing, yielding planet-to-star contrasts up to an order of magnitude better at $r$ $<$ 0\farcs{}5 than reported here.  This improved performance will reveal HD 36546's disk at even smaller separations and perhaps massive planets responsible for the disk's extreme luminosity.

\acknowledgements 
We thank Eric Mamajek for detailed discussions on HD 36546's age and Kevin Luhman, Scott Kenyon, Mengshu Xu, and the anonymous referee for other helpful comments.  We wish to emphasize the pivotal cultural role and reverence that the summit of Maunakea has always had within the indigenous Hawaiian community.  We are most fortunate to have the privilege to conduct scientific observations from this mountain.

{}

\begin{deluxetable}{lcccccccccc}
\setlength{\tabcolsep}{0pt}
\tablecolumns{4}
\tablecaption{Debris Disk Forward Modeling}
\tiny
\tablehead{{Parameter}&{Model Range}&{Best-Fit Model} & {Well-Fitting Models} }
\startdata
$i$(${\arcdeg}$) & 70 ... 80 & 75 & 70--75\\
$r_{\rm o}$ (au)&  75 ... 95 & 85& 75--95\\
$\alpha_{\rm in}$ & 3, 5, 10 & 3 & 3--10  \\
$\alpha_{\rm out}$ &$-1$, $-3$, $-5$, $-10$ & $-3$ & $-3$ \\
$g$ & 0 ... 0.85 & 0.85 & 0.7--0.85\\
$ksi_{\rm o}$ (au) & 5 ... 15& 10 & 5--15\\
\enddata
\label{diskmodels}
\end{deluxetable}

\begin{thebibliography}{}
\bibitem[Abt et al.(2004)]{Abt2004}Abt, H., 2004, \apjs, 155, 175
\bibitem[Augereau et al.(1999)]{Augereau1999}Augereau, J. C., Lagrange, A.-M., Mouillet, D., et al. 1999, A\&A, 348, 557
\bibitem[Baraffe et al.(2003)]{Baraffe2003}Baraffe, I., Chabrier, G., Barman, T. S., et al., 2003, A\&A, 402, 701
\bibitem[Brandt et al.(2014)]{Brandt2014}Brandt, T. D., Kuzuhara, M., McElwain, M. W., et al., 2014, \apj, 786, 1
\bibitem[Brandt and Huang(2015)]{BrandtHuang2015}Brandt, T. D., Huang, C., 2015, \apj, 807, 58
\bibitem[Carson et al.(2013)]{Carson2013}Carson, J., Thalmann, C., Janson, M., et al., 2013, \apj, 763, L32
\bibitem[Chini et al.(2012)]{Chini2012}Chini, R., Hoffmeister, V. H., Nasseri, A., et al., 2012, \mnras, 424, 1925
\bibitem[Choquet et al.(2015)]{Choquet2015}Choquet, E., Perrin, M. D., Chen, C. H., et al., 2016, \apj, 817, L2
\bibitem[Cloutier et al.(2014)]{Cloutier2014}Cloutier, R., Currie, T., Rieke, G. H., et al., 2014, \apj, 796, 127
\bibitem[Currie et al.(2008)]{Currie2008}Currie, T., Balog, Z., Kenyon, S. J., et al., 2008, \apj, 672, 558
\bibitem[Currie et al.(2011)]{Currie2011a}Currie, T., Burrows, A., Itoh, Y., et al., 2011, \apj, 729, 128
\bibitem[Currie et al.(2012)]{Currie2012}Currie, T., Debes, J., Rodigas, T., et al., 2012, \apj, 760, L32
\bibitem[Currie et al.(2014)]{Currie2014}Currie, T., Daemgen, S., Debes, J., et al., 2014, \apj, 780, L30
\bibitem[Currie et al.(2015)]{Currie2015a}Currie, T., Lisse, C., Kuchner, M., et al., 2015, \apj, 807, L7
\bibitem[Currie et al.(2016)]{Currie2016}Currie, T., Grady, C., Cloutier, R., et al., 2016, \apj, 819, L26
\bibitem[Garcia et al.(2016)]{Garcia2016}Garcia, E. V., Currie, T., Guyon, O., et al., 2016, \apj \ in press, arxiv:1610.05786
\bibitem[Groff et al.(2016)]{Groff2015}Groff, T. D., Chilcote, J., Kasdin, N. J., et al., 2016, SPIE, 9908, 0
\bibitem[Hernandez et al.(2004)]{Hernandez2004}Hernandez, J., Calvet, N., Briceno, C., et al., 2004, \aj, 127, 1682
\bibitem[Hog et al.(2000)]{Hog2000}Hog, E., Fabricius, C., Makarov, V. V., et al., 2000, A\&A, 355, 27
\bibitem[Jones et al.(2016)]{Jones2016}Jones, J., White, R. J., Quinn, S., et al., 2016, \apj, 822, L3
\bibitem[Jovanovic et al.(2015)]{Jovanovic2015}Jovanovic, N., Martinache, F., Guyon, O., et al., 2015, \pasp, 127, 890
\bibitem[Jovanovic et al.(2016)]{Jovanovic2016}Jovanovic, N., Guyon, O., Lozi, J., et al., 2016, SPIE, 9909, 0
\bibitem[Kenyon and Bromley(2008)]{KenyonBromley2008}Kenyon, S., Bromley, B., 2008, \apjs, 179, 451
\bibitem[Kenyon et al. (2008)]{Kenyon2008}Kenyon, S. J., Gomez, M., Whitney, B. A., 2008, Handbook of Star Forming Regions, Volume I: The Northern Sky ASP Monograph Publications, Vol. 4. Edited by Bo Reipurth, p.405
\bibitem[Lafreni\`ere et al.(2007)]{Lafreniere2007a}Lafreni\'ere, D., Marois, C., Duyon, R., et al., 2007, \apj, 660, 770
\bibitem[Lagrange et al.(2010)]{Lagrange2010}Lagrange, A.-M., Bonnefoy, M., Chauvin, G., et al., 2010, Science, 329, 57
\bibitem[Lagrange et al.(2012)]{Lagrange2012}Lagrange, A.-M., Boccaletti, A., Milli, J., et al., 2012, A\&A, 542, 40
\bibitem[Li and Hu(1998)]{Li98}Li, J. Z., Hu, J. Y., 1998, A\&AS, 132, 173
\bibitem[Luhman et al.(2009)]{Luhman2009}Luhman, K. L., Mamajek, E. E., Allen, P. R., Cruz, K., 2009, \apj, 703, 399
\bibitem[Macintosh et al.(2014)]{Macintosh2014}Macintosh, B., Graham, J., Ingraham, P., et al., 2014, PNAS, 111, 35
\bibitem[Mamajek(2016)]{Mamajek2016}Mamajek, E. E., 2016, https://dx.doi.org/10.6084/m9.figshare.3122689.v1
\bibitem[Marois et al.(2006)]{Marois2006}Marois, C., Lafreni\'ere, D., Duyon, R.,  al., 2006, \apj, 641, 556
\bibitem[Marois et al.(2008)]{Marois2008}Marois, C., Macintosh, B., Barman, T., et al., 2008, Science, 322, 1348
\bibitem[Marois et al.(2010)]{Marois2010b}Marois, C., Macintosh, B.,\& V\`{e}ran, J.-P., 2010, \procspie, 7736, 52
\bibitem[Mawet et al.(2014)]{Mawet2014}Mawet, D., Milli, J., Wahhaj, Z., et al., 2014, \apj, 792, 97
\bibitem[Murphy et al.(2013)]{Murphy2013}Murphy, S. J., Lawson, W., Bessell, M., 2013, \mnras, 435, 1325
\bibitem[Mustill and Wyatt(2009)]{Mustill2009}Mustill, A., Wyatt, M., \mnras, 399, 1403
\bibitem[Nesvold and Kuchner(2015)]{Nesvold2015}Nesvold, E., Kuchner, M., 2015, \apj, 798, 83
\bibitem[Pecaut et al.(2012)]{Pecaut2012}Pecaut, M., Mamajek, E., Bubar, E., 2012, \apj, 746, 154
\bibitem[Pecaut et al.(2013)]{Pecaut2013}Pecaut, M., Mamajek, E.,  2013, \apjs, 208, 9
\bibitem[Rodigas et al.(2014a)]{Rodigas2014}Rodigas, T. J., Malhotra, R., Hinz, P., 2014, \apj, 780, 65
\bibitem[Rodigas et al.(2014b)]{Rodigas2014b}Rodigas, T. J., Debes, J. H., Hinz, P., et al., 2014, \apj, 783, 21
\bibitem[Schneider et al.(2009)]{Schneider2009}Schneider, G., Weinberger, A. J., Becklin, E. E., et al., 2009, \aj, 137, 53
\bibitem[Slesnick et al.(2006a)]{Slesnick2006a}Slesnick, C., Carpenter, J. M., Hillenbrand, L. A., Mamajek, E. E., 2006a, \aj, 132, 2665
\bibitem[Slesnick et al.(2006b)]{Slesnick2006b}Slesnick, C., Carpenter, J. M., and Hillenbrand, L. A., 2006b, \aj, 131, 3016
\bibitem[Smith and Terrile(1984)]{SmithTerrile1984}Smith, B., Terrile, R., 1984, Science, 226, 1421
\bibitem[Thalmann et al.(2013)]{Thalmann2013}Thalmann, C., Janson, M., Buenzli, E., et al., 2013, \apj, 763, L29
\bibitem[Torres et al.(2007)]{Torres2007}Torres, R.M., Loinard, L., Mioduszewski, A.J, \& Rodriguez, L.F., 2007, \apj, 671, 1813
\bibitem[van Leeuwen(2007)]{vanLeewen2007}van Leeuwen, F., 2007, A\&A, 474, 653
\bibitem[Wu et al.(2013)]{Wu2013}Wu, C.-J., Wu, H., Lam, M.-I., et al., 2013, \apjs, 208, 29
\bibitem[Wyatt(2008)]{Wyatt2008}Wyatt, M. C., 2008, \araa, 46, 339

\end{thebibliography}
\end{document}